\documentclass[%
 preprint,prl,
showpacs,
 amsmath,amssymb,
 superscriptaddress,
 aps
]{revtex4-1}

\usepackage{amsmath}
\usepackage{amssymb}
\usepackage{graphicx}
\usepackage{dcolumn}
\usepackage{bm}
\usepackage{hyperref}
\usepackage{verbatim}
\usepackage{color}

\newcommand{\CH}[1]{\textcolor{black}{{#1}}}

\newcommand{\F}{F}  
\newcommand{\R}{\phi_R}  
\newcommand{\Rhom}{\bar{\phi}_R}  

\newcommand{\A}{\phi_A}  
\newcommand{\Ahom}{\bar{\phi}_A}  

\newcommand{\B}{\phi_B}  

\newcommand{\Out}{\tilde{\phi}_\text{out}} 

\begin{document}

\title{
\CH{Discontinuous switching of position of two coexisting phases}
}

\author{Samuel Kr\"uger}
\affiliation{Max Planck Institute for the Physics of Complex Systems,
N\"{o}thnitzer Str.~38, 01187 Dresden,
Germany
}
\affiliation{Leibniz-Institut f\"ur Polymerforschung, Dresden 01069, Germany}
\affiliation{Center for Advancing Electronics Dresden cfAED, Dresden, Germany}

\author{Christoph A.\ Weber}
\affiliation{Max Planck Institute for the Physics of Complex Systems,
N\"{o}thnitzer Str.~38, 01187 Dresden,
Germany
}
\affiliation{Center for Advancing Electronics Dresden cfAED, Dresden, Germany}
\affiliation{Division of Engineering and Applied Sciences, Harvard University, Cambridge, MA 02138, USA
}

\author{Jens-Uwe Sommer}
\affiliation{Leibniz-Institut f\"ur Polymerforschung, Dresden 01069, Germany}
\affiliation{TU Dresden, Institute for Theoretical Physics, Zellescher Weg 17, 01069 Dresden}
\affiliation{Center for Advancing Electronics Dresden cfAED, Dresden, Germany}

\author{Frank J\"ulicher}
\affiliation{Max Planck Institute for the Physics of Complex Systems,
N\"{o}thnitzer Str.~38, 01187 Dresden,
Germany
}
\affiliation{Center for Advancing Electronics Dresden cfAED, Dresden, Germany}

\begin{abstract}
Here we investigate 
how \CH{the positions of a condensed phase} can be controlled by using concentration gradients of a regulator that influences phase separation. 
We consider a mean field model of a ternary mixture where a  concentration gradient of  a regulator is imposed by an external 
 potential. We show that novel first order 
 phase transition exists at which \CH{the position of the condensed phase switches} in a discontinuous manner.  
\CH{This mechanism could have implications} for the spatial organization of biological cells and
provides a control mechanism for droplets in microfluidic systems.   
\end{abstract}

\maketitle

\CH{Phase separation of a mixture refers to the formation of a condensed phase that coexists with a dilute phase of lower concentration~\cite{Bray_Review_1994, Onuki_book}}. 
Such demixing is the result of a first order thermodynamic  phase transition
where \CH{the concentration difference between the phases changes discontinuously}.
It can be observed in many forms in everyday life, for example when oil is added to water. 
The occurrence of a transition from the homogeneous mixture to a system \CH{with coexisting phases} can be controlled  by temperature or by changing the composition of the mixture. \CH{Condensed phases are influenced by surfaces possibly causing wetting transitions~\cite{cahn1977critical,moldover1980interface,pohl1982wetting}.}
Furthermore, phase separation can be affected by external forces such as  gravity causing  sedimentation.

\textcolor{black}{A key question is how condensed phases such as droplets are positioned in systems with external cues like concentration gradients or external fields.}
The study of positioning of phases provides general insights
in the physics of phase separation of spatially inhomogeneous systems. 
\CH{Understanding the underlying mechanism of the positioning of condensed phases may open the possibility of applications in microfuidic devices.
Positioned condensed phases could be used to seal and open junctions  at specific locations in the microfluidic device, or simply position chemicals that enrich in the condensed or dilute phase.}
The positioning \CH{of condensed phases} in a complex mixture also plays a role in cell biology.
\CH{In particular, positioned condensed phases are used to
 segregate molecules during asymmetric cell division~\cite{brangwynne2011soft,hyman2014liquid,brangwynne2015polymer, saha2016polar}.}

\begin{figure}[t]
\centering
\includegraphics[width=1.0\textwidth]{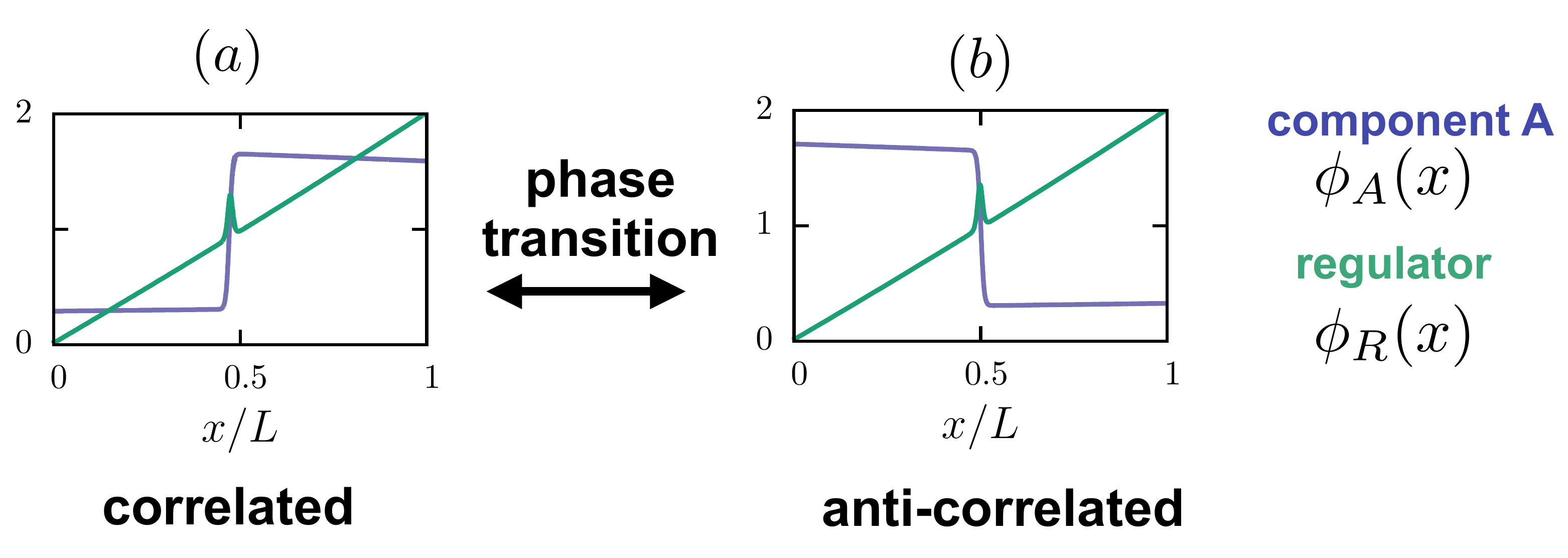}
\caption{\label{fig:Fig_0} 
Spatial regulation of phase separation by a discontinuous phase transition.
(a,b) The regulator (green) forms a gradient due to an external potential. 
Depending on the interactions with the regulator
the spatial distribution of e.g.\  component $A$ (purple; component $B$ behaves oppositely)
switches from a spatially correlated (a) to an anti-correlated (b) distribution with respect to the regulator. 
The switch corresponds to a discontinuous phase transition.
}
\end{figure}

\textcolor{black}{
Here we study the equilibrium physics of the positioning  
of two condensed phases in inhomogeneous systems.
We present a simplified model that provides the basic  mechanism for the positioning at thermal equlibrium
which can be further extended to non-equilibrium processes such as the kinetics of droplet formation and ripening.}
\textcolor{black}{In our model phase separation of two components is subject to a concentration gradient of a \emph{regulator component} where the gradient is generated by an external field.
The regulator component  affects demixing of the two components but does not phase separate itself.}
The system then relaxes to a spatially inhomogeneous thermodynamic equilibrium state
with two coexisting phases positioned by the regulator gradient.
\textcolor{black}{The spatial distributions of the three concentration profiles at thermal equilibrium are determined by minimizing a mean field free energy functional.}
We find that as a function of an interaction parameter the position of the condensed phase switches discontinuously from a position in the region of large regulator concentration (correlated state) to the region of low regulator concentration (anti-correlated). This switching of position corresponds to a novel, \CH{equilibrium} first order phase transition at which an order parameter undergoes a jump (Fig.~\ref{fig:Fig_0}(a,b)).

In our \CH{equilibrium} model for spatial regulation of phase separation
 we consider three components~\cite{Lee_2013}: two components which can demix from each other, $A$ and $B$, and a regulator $R$ that interacts with 
 these components. \CH{The regulator affects phase separation but does not demix from $A$ and $B$.}
 Demixing and interactions with the regulator are described 
by the Flory-Huggins free energy density for three components~(\cite{flory1942thermodynamics,huggins42} and Supplemental Material~\cite{SM}, II):
\begin{align}
 	\label{eq:free_energy_ternary_functional_theory}
	 f(\A,\R)  &= \frac{ k_bT}{\nu} \bigg[\sum_{i=A,B,R}\phi_i\ln\phi_i+ \chi_{AR}\, \A \R \\
	 \nonumber
	 &+\chi_{BR} \, \R \B+\chi_{AB} \,\A \B+(U/k_bT) \R \\
	 \nonumber
	 &+\frac{\kappa_R}{2}\left|\nabla\R\right|^2+\frac{\kappa_A}{2}\left|\nabla\A\right|^2+\frac{\kappa}{2}\nabla\R\nabla\A\bigg] \, .
\end{align}
We consider the incompressible system in which the molecular volumes are equal to $\nu$ and $\B=1-\R-\A$. 
The logarithmic contributions correspond to the mixing entropy, while the second line in Eq.~\eqref{eq:free_energy_ternary_functional_theory}  describes the molecular interactions between the components; $\chi_{ij}$ is the interaction parameter between component $i$ and $j$.
The gradient terms represent contributions to the free energy
associated with
spatial inhomogeneities. They introduce two  length scales, 
$\sqrt{\kappa_A}$ and  $\sqrt{\kappa_R}$.
The regulator $R$ is subject to an external field described by a position-dependent potential $U(x)$.
For simplicity we consider in the following
a one-dimensional system and choose
a potential of the form
$U(x)= - k_bT \ln{\left(1+s \left(2x-L \right) \right)}$, where $s>0$ characterizes the slope of the potential and its inverse is a 
third length scale in our model. 
Note that  in the absence of $A$ and for $\R \ll \B$, 
$\phi_R(x)$ attains a concentration profile that is linear in space with a slope $s$.
We consider a finite system of size $L$  and  two type of boundary conditions: (i)  Neumann boundary conditions,  $\phi_i' (0)= \phi_i' (L)=0$, for all fields, where the primes denote spatial derivatives, and (ii) periodic boundaries with $\phi_i (0)= \phi_i (L)$ and $\phi_i' (0)= \phi_i' (L)$. 
The conditions (i) imply that 
there is no explicit energetic bias to wet or dewet the boundary, but the presence of the boundary enforces the slopes of the concentration profiles close to the boundary. In contrast, the periodic conditions (ii) allow to study the system in the absence of  boundaries.

 To calculate the \CH{equilibrium profiles} $\A(x)$ and $\R(x)$, 
we minimize the  free energy 
\begin{equation}\label{eq:free_eneergy_tot}
	\F[\A(x),\R(x)] = \int_{0}^{L}\text{d}x \, f(\A(x),\R(x),x) \, .
\end{equation} 
Due to particle number conservation, 
two constraints are imposed for the minimization: Each field ($i=A,R$) obeys 
$\bar \phi_i = L^{-1} \int_0^L \text{d}x \, \phi_i(x)$, 
where  $\bar \phi_i$ are the average volume fractions and
$\bar \phi_B=1-\bar \phi_A-\bar \phi_R$.
Variation of the free energy Eq.~\eqref{eq:free_eneergy_tot} with the constraints of particle number conservation implies
($i=A,R$): 
\begin{equation}\label{eq:minimization_2}
	0=  \int_{0}^{L}\text{d}x \,  \left( \frac{\partial f}{\partial \phi_i} -
	 \frac{\text{d}}{\text{d}x} \frac{\partial{f}}{\partial\phi_i'}
 + \lambda_i \right) \delta \phi_i  + \frac{\partial f}{\partial \phi_i'} \delta \phi_i \bigg{|}_0^L\, ,
\end{equation}
where $\lambda_R$ and $\lambda_A$ are  Lagrange multipliers, and the prime denotes a derivative with respect to $x$ .
The boundary terms vanish for both, Neumann  and  periodic boundary conditions. 
Using the explicit form of the free energy density (Eq.~\eqref{eq:free_energy_ternary_functional_theory}), 
the Euler-Lagrange equations can be derived (see Supplemental Material~\cite{SM}, I)).
We solve these equations 
using a finite difference 
solver (bvp4c in MATLAB~\cite{Kierzenka_2001}).
As control parameters we consider the three interaction parameters 
$\chi_{AR}$, $\chi_{AB}$ and $\chi_{BR}$, the slope of the external potential $s$ and the mean volume fraction of $A$-material, $\bar \phi_A$.
The mean regulator material is fixed to $\bar \phi_R=0.02$ in all presented studies.
Moreover, we focus on the limit of strong phase segregation where the interfacial width is small compared to the system size, i.e.\ $\sqrt{\kappa_i} \ll L$.
In this limit, we verified that our results depend only weakly
on the specific values of $\kappa_i$.
%
\begin{figure}[t]
\centering
\begin{tabular}{cc}
\includegraphics[width=0.45\textwidth]{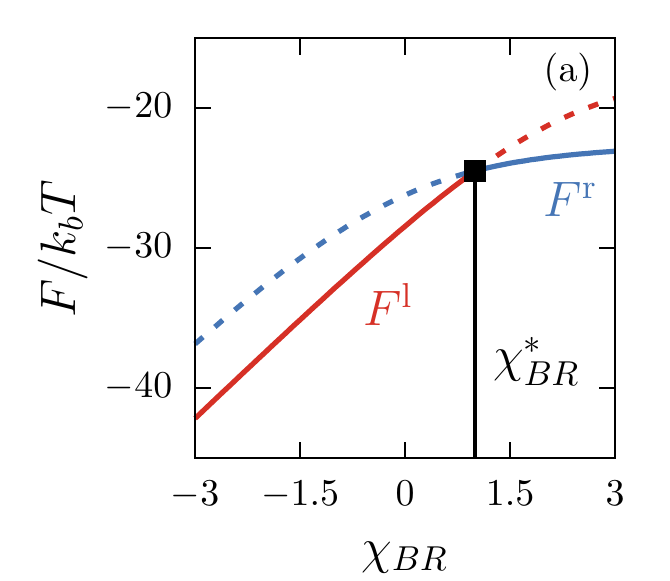} & 
\includegraphics[width=0.45\textwidth]{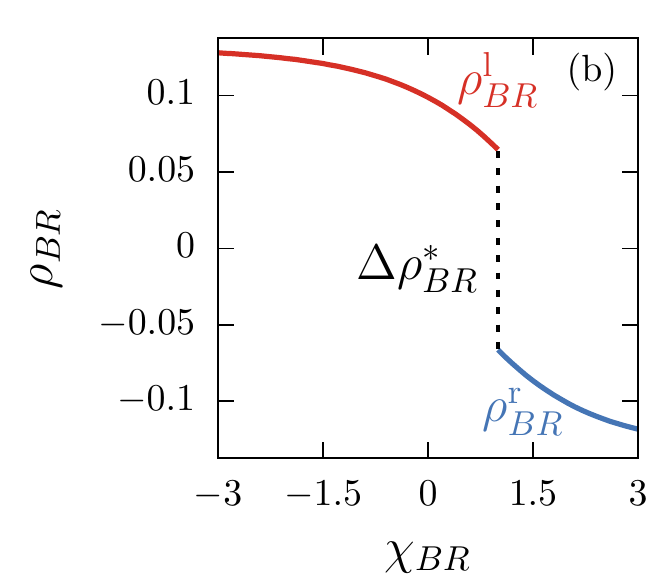} 
\end{tabular}
\caption{\label{fig:Fig_2} 
Discontinuous phase transition.
(a) Free energy $F$ (Eq.~\eqref{eq:free_eneergy_tot}) as a function of the $R$-$B$ interaction parameter $\chi_{BR}$.
$F^l$ and $F^r$ are the free energies of the correlated and anti-correlated 
stationary solution with respect to the regulator gradient, respectively (Fig.~\ref{fig:Fig_0}(a,b)). 
Lines are dashed when solutions are metastable. 
At $\chi_{BR}^*$, $F^l$ and $F^r$ intersect causing a kink 
 corresponding to the solution of lowest free energy. 
This shows that the transition between  correlation and anti-correlation
 is a discontinuous phase transition.
(b) The order parameter $\rho_{BR}$ (Eq.~\eqref{eq:order_parameter})
jumps at $\chi_{BR}^*$ by a value of $\Delta\rho_{BR}^*$.
The transition point $\chi_{BR}^*$ does not depend on the slope of the regulator $s$, while $\chi_{BR}^*$ increases linearly with $s$ (see Supplemental Material~\cite{SM}, VI)
Parameters: $\chi_{AB}=4$, $\chi_{AR}=1$, $\bar{\phi}_A=0.5$, $\bar{\phi}_R=0.02$, $\kappa_R/L^2=7.63 \cdot10^{-5}$, $\kappa_A/L^2=6.10\cdot10^{-5}$, $\kappa/L^2=6.10\cdot10^{-5}$, 
$Ls=0.99$. For plotting, $\nu=L/256$ was chosen.
}
\end{figure}
%
%

\newpage
\begin{figure*}[t]
\centering
\includegraphics[width=0.44\textwidth]{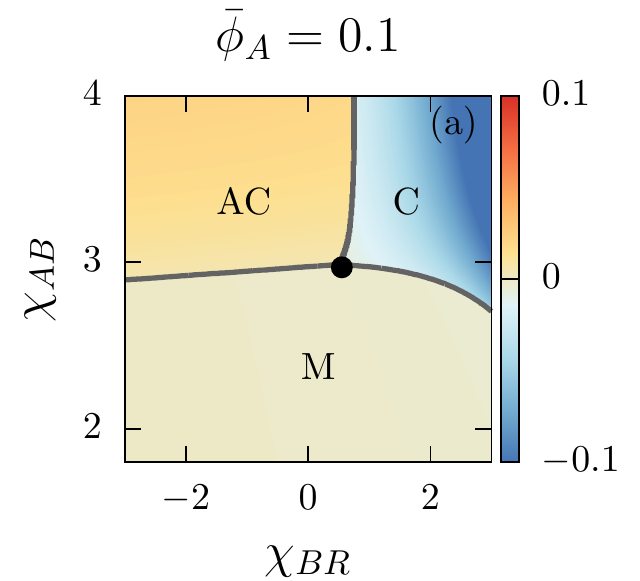} 
\includegraphics[width=0.44\textwidth]{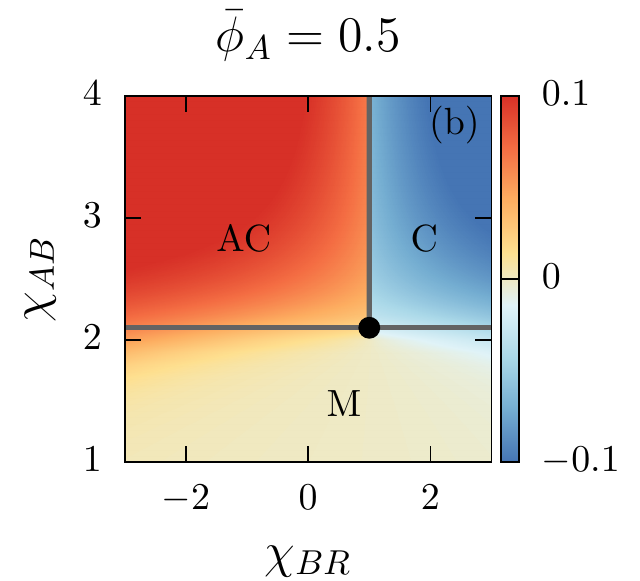}
\includegraphics[width=0.44\textwidth]{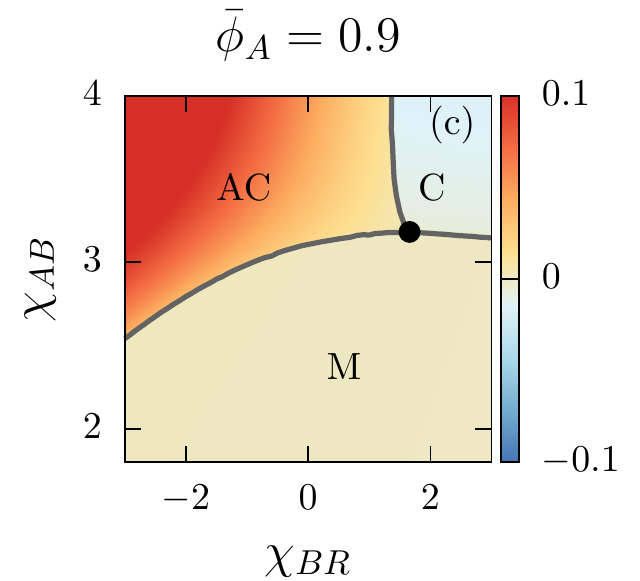}
\includegraphics[width=0.44\textwidth]{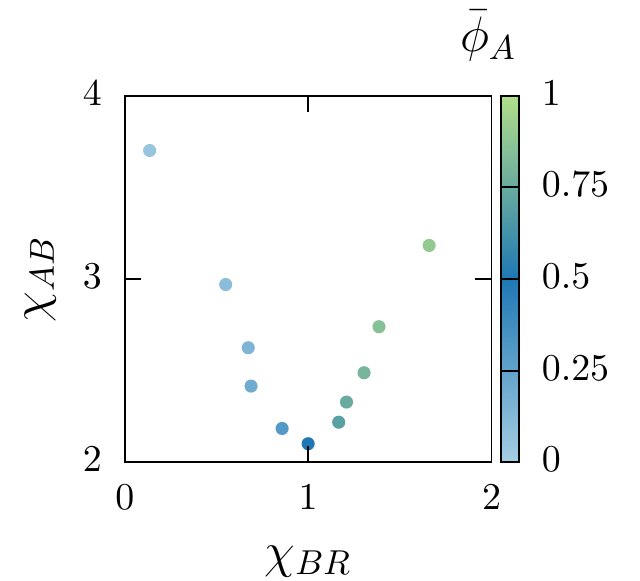} 
\caption{\label{fig:Fig_3}
Phase diagrams of our ternary model with spatial regulation. 
(a-c) 
Phase diagram for three volume fractions $\bar{\phi}_A=\{0.1, 0.5, 0.9 \}$ and varying the interaction parameters $\chi_{AB}$ and $\chi_{BR}$.
The color code depicts  the order parameter $\rho_{BR}$ defined in Eq.~\eqref{eq:order_parameter}.
Component $A$ is spatially correlated (C) with the regulator profile if 
$\rho_{BR}<0$, and anti-correlated (AC) otherwise. When the system is mixed (M),  $\rho_{BR} \approx 0$, and spatial profiles of all components are only weakly inhomogeneous (no phase separation).
 The solid black line in (a) is the transition line between C and AC calculated with the ansatz Eq.~\eqref{eq:analytic_sol} using condition~\eqref{eq:delta_F}.
The triple point (black dot) corresponds to the point in the phase diagrams where the three regions meet and the three free energies are equal.
 (d) Triple point for different  $\bar{\phi}_A$ values (color code). 
Parameters:  $\chi_{AR}=1$, $\bar{\phi}_R=0.02$, $\kappa_R/L^2=7.63 \cdot10^{-5}$, $\kappa_A/L^2=6.10\cdot10^{-5}$, $\kappa/L^2=6.10\cdot10^{-5}$, $Ls=0.99$, $\nu=L/256$. 
}
\end{figure*}
\clearpage

Solving the Euler-Lagrange equations with Neumann boundary conditions (i), we find two 
spatially inhomogeneous solutions  for component $A$, 
which we denote  $\A^\text{l}(x)$ and  $\A^\text{r}(x)$,
and the two corresponding solutions for the regulator component $R$, 
are denoted $\R^\text{l}(x)$ and $\R^\text{r}(x)$ 
(the profile of $B$ follows from volume conservation). 
 The phase separating material $A$ is either accumulated 
 close to the right boundary of the system ($\R^\text{r}(x)$ and $\A^\text{r}(x)$) and correlated with the 
concentration of the regulator material (Fig.~\ref{fig:Fig_0}(a)),
or it is accumulated at the left ($\R^\text{l}(x)$ and $\A^\text{l}(x)$)  and anti-correlated with the regulator (Fig.~\ref{fig:Fig_0}(b)).
Upon varying the interaction parameters $\chi_{BR}$ in Fig.~\ref{fig:Fig_2}(a,b), the free energies of the correlated and the anti-correlated states, $\F^\text{r}=\F[\A^\text{r}, \R^\text{r}]$ and $\F^\text{l}=\F[\A^\text{l}, \R^\text{l}]$,  are 
different. They intersect at one point $\chi_{BR}=\chi_{BR}^*$ (Fig.~\ref{fig:Fig_2}(a)).
At this point the lowest free energy exhibits a kink, which 
 means that the system undergoes a discontinuous phase transition when switching from the 
 spatially anti-correlated (`left') to the spatially correlated (`right') solution with respect to the regulator.
A set of order parameters suitable to study this phase transition is 
\begin{align}
\nonumber
	\rho_{ij} &= \left( k_bT L \mathcal{N}_{ij} /\nu\right)^{-1} 
	\frac{\text{d}}{\text{d}\chi_{ij}} \left[ F(\phi_i(x),\phi_j(x)) - F(\bar \phi_i, \bar \phi_j) \right] \\
	\label{eq:order_parameter}
	 &=\mathcal{N}^{-1}_{ij} \int_0^L \text{d}x \,  \left( \phi_i(x) \phi_j(x) - \bar \phi_i \bar \phi_j\right) \, ,
\end{align} 
where the squared normalization $\mathcal{N}^2_{ij}= \text{Var}( \phi^{\Theta}_{i})   \text{Var}( \phi^{\Theta}_{j})$ with  $\text{Var}(\phi_i)=  \int_0^L \text{d}x \,  \left( \phi_i^2(x) - \bar \phi_i^2 \right)$, denoting the variance and $\phi^{\Theta}_i(x)=\Theta{(L\bar \phi_i -x)}$, where $\Theta(\cdot)$ denotes the Heaviside step function.
 This normalization ensures that $-1<\rho_{ij}<1$ and
 $\rho_{ij}=\pm1$ if $\phi_i(x)=\phi^{\Theta}_i(x)$.
The derivative of the free energy with respect to the interaction parameter $\chi_{ij}$ 
generates the covariance between the spatially dependent fields $\phi_i(x)$ and $\phi_j(x)$.
If the fields are spatially correlated, $\rho_{ij}>0$, and if they are anti-correlated, $\rho_{ij}<0$. For homogeneous fields with $\phi_i(x)=\bar \phi_i$, $\rho_{ij}=0$.
Varying the interaction parameter $\chi_{BR}$  (Fig.~\ref{fig:Fig_2}(b)), the order parameters $\rho_{BR}$ and  $\rho_{AR}$ jump at the threshold value $\chi_{BR}^*$, while in the absence of a regulator gradient ($s=0$), they change smoothly. 
 The jump of both order parameters in the presence of a regulator gradient indicates that the spatial correlation of $A$ and $B$ to $R$ changes abruptly, which is expected in case of a first order phase transition.

By means of the order parameter 
$\rho_{BR}$ (Eq.~\eqref{eq:order_parameter}) we can now discuss the phase diagrams as a function of the interaction parameters 
  for different volume fractions of the demixing material, $\Ahom$.
 \textcolor{black}{We find three regions (Fig.~\ref{fig:Fig_3}(a-c)): 
 A mixed region (M), where volume fraction profiles are only weakly inhomogeneous and no phase separation occur. 
  In addition, there are two regions, (C) and (AC), where components $A$ and $B$ 
  phase separate and $A$
  is  spatially correlated or anti-correlated with the regulator $R$, respectively.}
  There exists a triple point where all three states have the same free-energy. 
\textcolor{black}{For $\Ahom=1/2$, the shape of the transition line between correlated and anti-correlated states is straight and $\chi_{BR}^*$ is independent of $\chi_{AB}$  (Fig.~\ref{fig:Fig_3}(b)).}
\textcolor{black}{If $\Ahom$ is decreased, the region of the correlated state in the phase diagram grows.
In this case, the correlated state is favored,
while for increasing  $\Ahom$, the anti-correlated state is preferred.}
 The transition line to the mixed states is horizontal for $\Ahom=1/2$
    (Fig.~\ref{fig:Fig_3}(b)). 
     For both, larger and smaller  $\Ahom$-values, it 
    becomes curved and moves towards larger $\chi_{AB}$ interaction parameters. 
    This behavior can be qualitatively understood by the upshift of the demixing threshold $\chi_{AB}$ once $\Ahom$ deviates from $1/2$,  as known for binary systems. Since the concentration of $R$ is
    small here, this analogy provides a good approximation
    ($\bar \phi_R\to 0$ in Eq.~\eqref{eq:free_energy_ternary_functional_theory}).
  Both trends explain the parabolic shape of the positions of the triple point in the phase diagrams when $\bar\phi_A$ is varied
    (Fig.~\ref{fig:Fig_3}(d)).

The transition line in the phase diagrams between the correlated and anti-correlated solution as a function of the interaction parameters
 can be estimated analytically.   
  In the absence of a regulator gradient ($s=0$), the free energies of both solutions are the same  for all interaction parameters for which phase separation occurs.
 In the presence of a regulator gradient, however, the free energies corresponding 
 to the correlated and the anti-correlated solutions are unequal for most points in the phase diagram. The reason is that the external potential 
  $U(x)$ forces the regulator to form a gradient, and thus the interactions with the regulator lead to different free energies of the correlated and anti-correlated states.
Only along the transition line between both states  the free energies equal:
 \begin{equation}\label{eq:delta_F}
	 \Delta \F=\F[\A^\text{r}, \R^\text{r}]-\F[\A^\text{l}, \R^\text{l}]=0 \, .
 \end{equation}
 \CH{This condition can be used to estimate the transition line for varying  interaction parameters and the slope of the potential, $s$.}
To estimate $ \Delta \F$
we parametrize the profiles of the  stationary solutions $\A^\text{r,l}(x)$ and $\R^\text{r,l}(x)$ using physical assumptions that are in  agreement with our numerical results.
First we idealize the already  narrow interface of the demixed component $\A$ as sharp. Since the regulator is maintained by the external potential, we find $\R^\text{r}(x)\simeq \R^\text{l}(x)$ close to the transition line. 
Thus we use the one profile, denoted as $\R(x)$, for both regulator states.
In addition, we approximate the regulator profile as linear function with slope $m$, 
neglecting spatial non-linearities that can be seen in Fig.~\ref{fig:Fig_0}(a,b).
The low volume fractions outside the condensed phase 
of the demixed binary $A$-$B$ system
 are approximated as constant values $\tilde \phi_\text{out}$.
The larger volume fraction (inside) shows a weakly  
linear profile (Fig.~\ref{fig:Fig_0}(a,b)). 
For most parameters, the volume fraction inside the condensed phase can be 
well described as $\phi_\text{in}(x)=\tilde \phi_\text{in}-\phi_R(x)$,
where $\tilde \phi_\text{in}$ is the constant volume fraction 
inside the condensed phase of the binary $A$-$B$ mixture (see Supplemental Material~\cite{SM}, V). 
The approximated profiles are:
 \begin{subequations}\label{eq:analytic_sol}
\begin{align}
\A^\text{l}(x)&= \left[\phi_\text{in}(x)-\tilde \phi_\text{out}\right] \, \Theta{(\epsilon_\text{l}-x)}  +\tilde \phi_\text{out} \, , \\
\A^\text{r}(x)&= \left[\phi_\text{in}(x)-\tilde\phi_\text{out} \right] \, \Theta{(\epsilon_\text{r}-L+x)}  +\tilde \phi_\text{out} \, , \\
\R(x)&= m  \left( x -L/2 \right) + \bar \phi_R \, .
\end{align}
 \end{subequations}
The conservation of $A$
determines  the domain sizes $\epsilon_\text{l,r}$ of the  phase separated region  (see Supplemental Material~\cite{SM}, IV).
\begin{figure}[t]
\centering
\begin{tabular}{cc}
\includegraphics[width=0.44\textwidth]{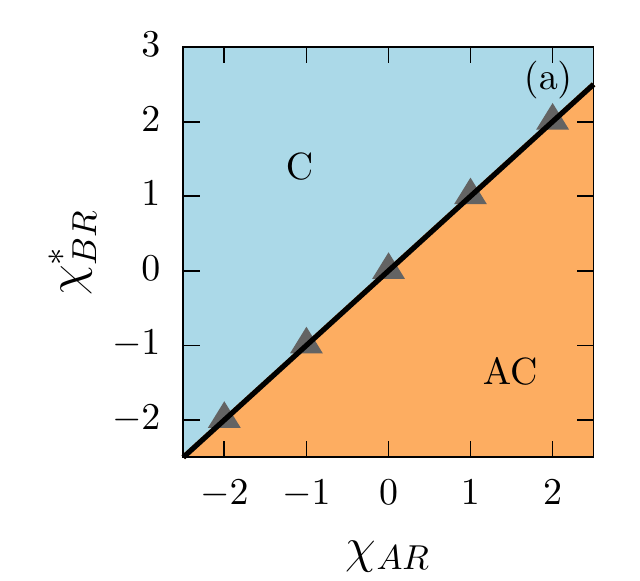} & 
\includegraphics[width=0.44\textwidth]{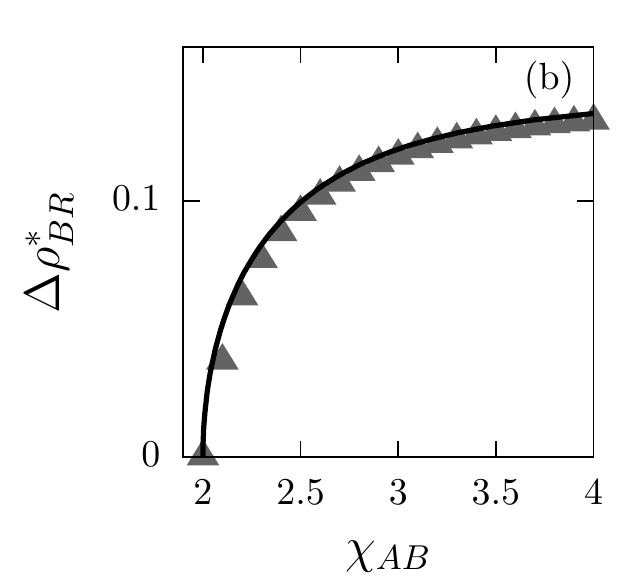} 
\end{tabular}
\caption{\label{fig:Fig_4}
Phase diagrams and order parameters estimated by ansatz Eq.~\eqref{eq:analytic_sol}. 
(a) The transition between spatial correlation (C) and anti-correlation (AC) of the distribution of  component $A$ with respect to the regulator in the $\chi_{AR}$-$\chi_{BR}$-plane. 
Parameters:  $\chi_{AB}=4$, $\bar{\phi}_A=0.5$, $\bar{\phi}_R=0.02$, $\kappa_R/L^2=7.63 \cdot10^{-5}$, $\kappa_A/L^2=6.10 \cdot10^{-5}$, $\kappa/L^2=6.10\cdot10^{-5}$, $Ls=0.99$, $\nu=L/256$.
(b) Jump of the order parameter at the transition point, $\Delta\rho_{BR}^*$, as a function of the interaction parameter $\chi_{AB}$. 
 Additionally to the parameters of (a), $\chi_{AR}=1$ and $\chi_{BR}=1$. 
The black line in (a) and (b) shows the result obtained from using Eq.~\eqref{eq:analytic_sol}; the triangles are numerical results from the minimization of Eq.~\eqref{eq:free_eneergy_tot}. 
}
\end{figure}
To calculate  $\Delta \F$  (Eq.~\eqref{eq:delta_F}),  the free energy density (Eq.~\eqref{eq:free_energy_ternary_functional_theory}) is integrated in the domain $[0,L]$. Using the approximated 
profiles (Eqs.~\eqref{eq:analytic_sol}) we find 
\begin{align}
	\Delta \F &\simeq \frac{k_bT}{\nu}\frac{\chi_{BR}-\chi_{AR}}{12} m 
	\, \mathcal{G}\, ,
\end{align}
where the value $\mathcal{G}$ depends only on the parameters of the simplified solutions (see Supplemental Material~\cite{SM}, IV). 
Consistently, $\Delta \F=0$, if there is no regulator gradient ($m=0$).
In presence of a regulator gradient,
$\Delta \F=0$ if 
$\chi_{BR}^* = \chi_{AR}$, 
which defines the transition line between the correlated and anti-correlated solution
obtained from the parametrized solutions Eqs.~\eqref{eq:analytic_sol}.
 This prediction is in very good agreement with our numerical results for $\bar \phi_A\simeq 1/2$;  see black lines in Fig.~\ref{fig:Fig_3}(a) and Fig.~\ref{fig:Fig_4}(a).
By means of the ansatz given in Eqs.~\eqref{eq:analytic_sol}
we can also  estimate how the jump of the order parameter 
 $\Delta\rho_{BR}^*$ (definition see Fig.~\ref{fig:Fig_2}(b))  at the transition point  depends on the model parameters.
  In particular we find that the estimated $\Delta\rho_{BR}^*$  as a function of the slope of the regulator (not shown) and the interaction parameter $\chi_{AB}$ (Fig.~\ref{fig:Fig_4}(b))  almost perfectly describe the data obtained from  the numerical minimization of the free energy.
This shows that the proposed parametrization of the stationary solutions represents a consistent approximation.
We conclude that the positioned and phase separated  profiles possess a sharp interface and the volume fraction inside has a weak linear slope that is mainly determined by volume exclusion with the regulator.

 The phase diagrams (Fig.~\ref{fig:Fig_3}) depend on the boundary conditions
 rising the question whether the boundary play a key role for the  existence of the phase transition. 
  To this end we considered a periodic system without boundaries. 
 We find that the reported first order transition  also exists for in the absence of boundaries (see Supplemental Material~\cite{SM}, III).
  Thus the transition is not induced by boundaries as for example in the case of wetting transitions~\cite{cahn1977critical,moldover1980interface,pohl1982wetting}.



  \CH{The discontinuous switching of phase separation  could be tested experimentally.
 A  soluble salt of high magnetic susceptibility could be used to create and maintain concentration gradients via the application of an inhomogeneous  magnetic field~\cite{Particle_gradient_due_to_magnetic_field}. 
Phase separation in a regulator gradient could be observed 
by introducing components that phase separate in a salt dependent manner. 
In particular, a pre-formed droplet could be added to an existing regulator gradient or 
the regulator gradient is created after Ostwald-ripening is completed~\cite{Lifshitz_Slyozov_61,wagner61,yao1993theory,weber2017droplet}. 
The phase transition could be triggered by 
 changing the concentrations of the phase separating material, by changing the temperature or by adding additional components that influence the interaction parameters.}
 The systems considered here could also be relevant for applications.
\CH{As the composition of a  condensed phase creates a distinct chemical environment, our work may provide a novel mechanism
 to control and switch chemical environments 
in microfluidic devices.}

 \newpage
 
\begin{acknowledgments}
We would like to thank Martin Elstner and Omar Adame for fruitful and stimulating discussions. This project was supported by the Center for Advancing Electronics Dresden (cfAED). 
Christoph A.\ Weber thanks the German Research Foundation (DFG) for financial support. Samuel Kr\"uger and Christoph A.\ Weber contributed equally to this work.
\end{acknowledgments}


\newpage
\clearpage

\section*{Supplemental Material}

\subsection{Euler-Lagrange Equations}

From the variation of the free energy and the explicit form of the free energy density (main text, Eq.(1)), we find the following set of the Euler-Lagrange equations:
\begin{subequations}
\label{eq:Euler_Lagrange_Eq}
\begin{align}
0&=\left(\kappa_A\kappa_R-\frac{\kappa^2}{4}\right)\A^{\prime \prime}-\kappa_R\bigg(\lambda_A+\chi_{AB}\left(1-2\phi_A\right)+\chi\phi_R\nonumber\\
&+\ln\left(\frac{\phi_A}{1-\phi_A-\phi_R}\right)\bigg)+\frac{\kappa}{2}\bigg(\lambda_R+U_R+\chi_{BR}\left(1-2\phi_R\right)\nonumber\\
&+\chi\phi_A+\ln\left(\frac{\phi_R}{1-\phi_A-\phi_R}\right)\bigg),\\
0&=\left(\kappa_A\kappa_R-\frac{\kappa^2}{4}\right)\R^{\prime \prime}-\kappa_A\bigg(\lambda_R+U_R+\chi_{BR}\left(1-2\phi_R\right)\nonumber\\
&+\chi\phi_A+\ln\left(\frac{\phi_R}{1-\phi_A-\phi_R}\right)\bigg)+\frac{\kappa}{2}\bigg(\lambda_A+\chi_{AB}\left(1-2\phi_A\right)\nonumber\\
&+\chi\phi_R+\ln\left(\frac{\phi_A}{1-\phi_A-\phi_R}\right)\bigg).
\end{align}
\end{subequations}
Here, we defined $\chi=\chi_{AR}-\chi_{AB}-\chi_{BR}$
and rescaled length $x\to x \,  L$.

\subsection{Penalty of spatial inhomogeneities in the ternary Flory-Huggins free energy density}

\subsubsection{Derivation using a mean field approximation}
\label{eq:mean_field}

To show this relation, we start from the local mean-field free energy on the lattice and calculate the continuum limit of this free energy as shown in Ref.~\cite{safran_book} for a binary system.
The local free energy density of the three component system is derived in \cite{Sivardiere_II_1975, Sivardiere_III_1975} using a mean-field approximation:
\begin{align}
 \frac{f\nu}{k_\text{B}T} = &\sum_{\alpha}\left(\phi_A(\alpha)\ln\phi_A(\alpha) + \phi_R(\alpha)\ln\phi_R(\alpha) + \left(1 - \phi_A(\alpha) - \phi_R(\alpha)\right)\ln\left(1 - \phi_A(\alpha) - \phi_R(\alpha)\right)\right) \nonumber\\
&+\frac{1}{2}\sum_{\alpha,\beta\text{ with }\alpha\ne \beta}\left(J_{AB}(\alpha,\beta)\phi_A(\alpha)\left(1 - \phi_A(\beta) - \phi_R(\beta)\right)\right. \nonumber\\
&\left.+ J_{BR}(\alpha, \beta)\phi_R(\alpha)\left(1 - \phi_A(\beta) - \phi_R(\beta)\right) + J_{AR}(\alpha, \beta)\phi_A(\alpha)\phi_R(\beta)\right) \, ,
\end{align}
where $\nu$ is the molecular volume.
The greek indices $\alpha$ and $\beta$ indicate the positions on the lattice.
The first line describes the entropy of the mixture.
Each contribution is local. 
The second and third line contains the energetic part of the free energy. 
It describes the non-local interactions between neighboring lattice sites. 

In the next steps we will perform the continuum limit. 
In case of the entropic contribution, we can simply replace $\phi_i(\alpha) \to \phi_i(x)$. In case of the energetic contributions, we rearrange the terms leading to:
\begin{align}
\frac{1}{2}\sum_{\alpha,\beta\text{ with }\alpha\ne \beta}\left[J_{AB}(\alpha,\beta)\phi_A(\alpha)\left(1 - \phi_A(\beta)\right) + J_{BR}(\alpha,\beta)\phi_R(\alpha)\left(1 - \phi_R(\beta)\right) \right.\nonumber\\
\left.+ \left(J_{AR}(\alpha,\beta) - J_{AB}(\alpha,\beta) - J_{BR}(\alpha,\beta)\right)\phi_A(\alpha)\phi_R(\beta)\right].
\end{align}
Each contribution can be rewritten as
\begin{align}
&J_{AB}(\alpha,\beta)\phi_A(\alpha)\left(1 - \phi_A(\beta)\right) \\
\nonumber
&\quad = \frac{1}{2}J_{AB}(\alpha,\beta)\left[\left(\phi_A(\alpha) - \phi_A(\beta)\right)^2 - \left(\phi_A(\alpha)\right)^2 - \left(\phi_A(\beta)\right)^2 + 2\phi_A(\alpha)\right] \, ,\\
&J_{BR}(\alpha,\beta)\phi_R(\alpha)\left(1 - \phi_R(\beta)\right) \\
\nonumber
& \quad= \frac{1}{2}J_{BR}(\alpha,\beta)\left[\left(\phi_R(\alpha) - \phi_R(\beta)\right)^2 - \left(\phi_R(\alpha)\right)^2 - \left(\phi_R(\beta)\right)^2 + 2\phi_R(\alpha)\right] \, ,\\
&\left[J_{AR}(\alpha,\beta) - J_{AB}(\alpha,\beta) - J_{BR}(\alpha,\beta)\right]\phi_A(\alpha)\phi_R(\beta) \nonumber\\
&\quad = \frac{1}{2}\left[J_{AR}(\alpha,\beta) - J_{AB}(\alpha,\beta) - J_{BR}(\alpha,\beta)\right]\left[\phi_A(\alpha)\phi_R(\alpha) + \phi_A(\beta)\phi_R(\beta) \right.\nonumber\\
&\quad \left. - \left(\phi_A(\alpha) - \phi_A(\beta)\right)\left(\phi_R(\alpha) - \phi_R(\beta)\right)\right].
\end{align}
%
We can identify the Flory Huggins interaction parameter as $\chi_{ij}=\frac{1}{2}\sum_{\beta}J_{ij}(\alpha,\beta)$.
In the continuum limit we can introduce the gradient of the volume fractions as $\left(\phi_i(\alpha) - \phi_i(\beta)\right)\rightarrow a\nabla\phi_i$.
We finally obtain the free energy
$F = \int dx\, f$ with the free energy density given as
\begin{align}
f = f_0\left(x\right) + \frac{k_\text{B}T}{\nu} \left[\frac{\kappa_A}{2}|\nabla\phi_A\left(x\right)|^2 + \frac{\kappa_R}{2}|\nabla\phi_R\left(x\right)|^2 + \frac{\kappa}{2}\nabla\phi_A\left(x\right)\nabla\phi_R\left(x\right) \right] \, ,
\end{align}
where 
\begin{align}
 \frac{f_0\nu}{k_\text{B}T} &= \phi_A\left(x\right)\ln\phi_A\left(x\right) + \phi_R\left(x\right)\ln\phi_R\left(x\right) + \left(1 - \phi_A\left(x\right) - \phi_R\left(x\right)\right)\ln\left(1 - \phi_A\left(x\right) - \phi_R\left(x\right)\right) \nonumber\\
& + \chi_{AR}\phi_A\left(x\right)\phi_R\left(x\right) + \chi_{AB}\phi_A\left(x\right)\left(1 - \phi_A\left(x\right) - \phi_R\left(x\right)\right) + \chi_{BR}\phi_R\left(x\right)\left(1 - \phi_A\left(x\right) - \phi_R\left(x\right)\right) \, .
\end{align}
The parameters characterizing the penalty corresponding to spatial inhomogeneities are $\kappa_{i}=a^2\chi_{i B}$, $i\in \{A,R \}$, and $\kappa = a^2\left(\chi_{AR} - \chi_{AB} - \chi_{BR}\right)$.

\subsubsection{Phenomenological derivation}

In the  Ginzburg-Landau free energy
the penalties corresponding to spatial inhomogeneities
are phenomenologically introduced based on symmetry considerations:
\begin{equation}
f - f_0 = \frac{\tilde \kappa_A}{2}\left(\nabla\phi_A\right)^2 + \frac{\tilde \kappa_B}{2}\left(\nabla\phi_B\right)^2 + \frac{\tilde \kappa_R}{2}\left(\nabla\phi_R\right)^2\, ,
\end{equation}
where $\tilde \kappa_i>0$ since spatial inhomogeneities are unfavored. Moreover, $f_0$ is the free energy density that only depends on the volume fractions $\phi_i$, $i\in A,B,R$.
However, only two volume fraction fields are independent due to particle conservation and incompressibility, $1=\phi_A+\phi_B+\phi_R$.
Thus we can write $\nabla\phi_B = -\nabla\phi_A - \nabla\phi_R$, leading to
\begin{equation}
f - f_0 = \frac{\kappa_A}{2} \left(\nabla\phi_A\right)^2 
+ \frac{\kappa_R}{2}\left(\nabla\phi_R\right)^2 + \frac{\kappa}{2}\nabla\phi_A\nabla\phi_R \, .
\end{equation}
Here, $ \kappa_A=\tilde \kappa_A + \tilde \kappa_B$,
$\kappa_R=\tilde \kappa_R + \tilde \kappa_B$ and $ \kappa=\tilde \kappa_B$.

\subsubsection{Choice of the parameters $\kappa_i$}

In the presented studies, we have chosen $\kappa_A=\kappa$ for simplicity. Please note that the derivation presented in Sect.~\ref{eq:mean_field} 
is based on a mean field approximation and therefore it should only serve as an estimate for the values $\kappa_i$. We chose  the values for the  parameters $\kappa_A$ and $\kappa_R$ consistent with these estimates  (see figure captions in the main text).

\subsection{Discontinuous phase transition in a periodic domain}

\begin{figure}[tbh]
\centering
\begin{tabular}{ccc}
\includegraphics[width=0.3\textwidth]{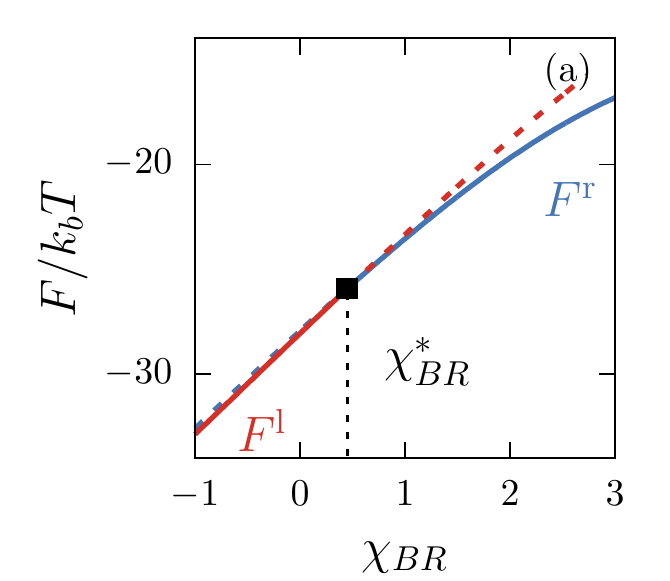} & 
\includegraphics[width=0.3\textwidth]{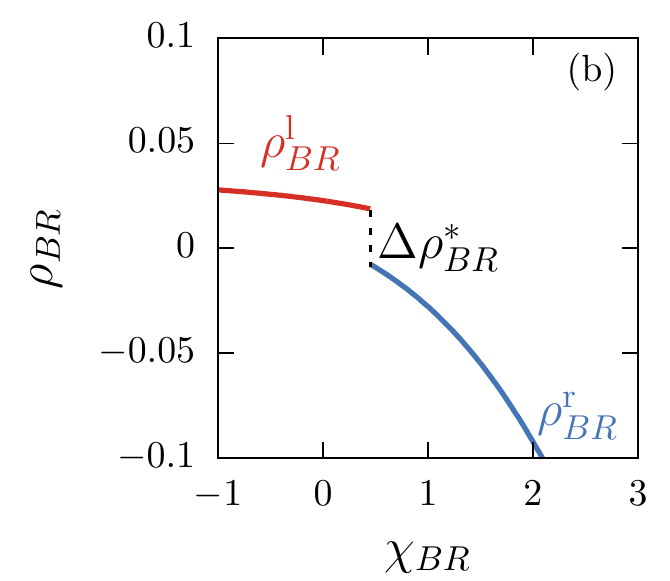} &
\includegraphics[width=0.33\textwidth]{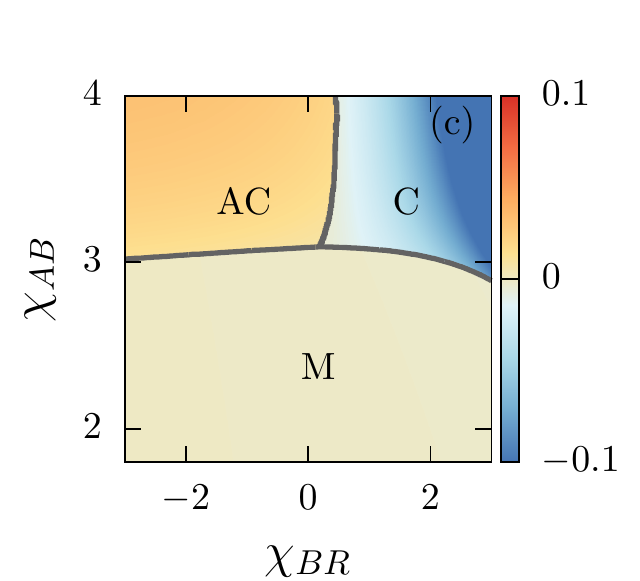} 
\end{tabular}
\caption{\label{fig:Fig_2SM} 
Discontinuous phase transition in a periodic potential and periodic boundary conditions.
(a) Free energy $F$ as a function of the $B$-$R$ interaction parameter $\chi_{BR}$.
$F^l$ and $F^r$ are the free energies of the correlated and anti-correlated 
stationary solution with respect to the regulator gradient, respectively. 
Lines are dashed when solutions are metastable. 
At $\chi_{BR}^*$, $F^l$ and $F^r$ intersect and the solution of lowest free energy exhibits a kink. 
This shows that the transition between  correlation and anti-correlation
 is a discontinuous phase transition.
(b) The order parameter $\rho^{RB}$
jumps at $\chi_{BR}^*$ by a value of $\Delta\rho_{BR}^*$.
Parameters: $\chi_{AB}=4$, $\chi_{AR}=1$, $\bar{\phi}_A=0.1$, $\bar{\phi}_R=0.02$, $\kappa_R/L^2=7.63 \cdot10^{-5}$, $\kappa_A/L^2=6.10\cdot10^{-5}$, $\kappa/L^2=6.10 \cdot10^{-5}$,
$A=0.5$. For plotting, $\nu=L/256$ was chosen.
(c) Phase diagrams of our ternary model for spatial regulation in a periodic potential and periodic boundary conditions ($\bar{\phi}_A=0.1$). 
The color code depicts  the order parameter $\rho_{BR}$.
Component $A$ is spatially correlated (C) with the regulator profile if 
$\rho_{BR}<0$, and anti-correlated (AC) otherwise. When the system is mixed (M),  $\rho_{BR} \approx 0$, and spatial profiles of all components are only weakly inhomogeneous (no phase separation).
The triple point (black dot) corresponds to the point in the phase diagrams where the three regions meet and the three free energies are equal.
Parameters:  $\chi_{AR}=1$, $\bar{\phi}_A=0.1$, $\bar{\phi}_R=0.02$, $\kappa_R/L^2=7.63 \cdot10^{-5}$, $A=0.5$, $\nu=L/256$. 
}
\end{figure}

Here we discuss the results of the minimization of the free-energy (Eq.~(3), main text) using
periodic boundaries with $\phi_i (0)= \phi_i (L)$ and $\phi_i' (0)= \phi_i' (L)$. 
We find the same main results as for Neumann boundary conditions, namely the existence of a discontinuous phase transition.
In the periodic domain, we also use a periodic external potential:
\begin{equation}
U=-k_bT\ln\left(1-A\sin\left(2\pi\left(\frac{x}{L}-\omega\right)\right)\right).
\end{equation} 
The parameter $\omega$ is a phase shift.
The value of the phase is chosen such that that the region of segregated A-material is placed at $x=0$.
The logarithmic form of the potential is chosen ensures that a sinus distribution of the regulator is obtained in the dilute limit.
We find two stationary solutions of different spatial correlations with respect to the regulator.
They switch at $\chi_{BR}^*$ by a discontinuous phase transition (Fig.~\ref{fig:Fig_2SM}(a-c)).
Therefore, a boundary of the system is not a necessary requirement  for the emergence of the discontinuous phase transition discussed in our manuscript.

%


\clearpage
\subsection{Estimate of $\Delta \F$}

The free energy difference between the two stationary solutions,  $\Delta \F$, results from integration over the domain $[0,L]$ using the simplified solutions 
(Eqs.~(7), main text): 
\begin{align}
	\Delta \F &= \frac{k_bT}{\nu}\frac{\chi_{BR}-\chi_{AR}}{12} m 
	\, \mathcal{G}\, ,
\end{align}
where 
\begin{equation}
\mathcal{G}=	\bigg[12 \epsilon_\text{l} \left( L-\epsilon_\text{l}\right) \left(2\Out+2\Rhom-1\right)+3 \left(L -2\epsilon_\text{l} \right) \left( m\left(L-2\epsilon_\text{l}\right)+4\Out+2\Rhom-2 \right) \Delta\epsilon\bigg]
\end{equation}
and
\begin{align}
\epsilon_\text{l}=&\frac{-2-Lm+4\Out+2\Rhom}{2m} +\frac{\sqrt{8Lm\left(\Out-\Ahom\right)+\left(2+Lm-4\Out-2\Rhom\right)^2}}{2m} \, ,\\
\epsilon_\text{r}=&\frac{-2+Lm+4\Out+2\Rhom}{2m} +\frac{\sqrt{8Lm\left(\Ahom-\Out\right)+\left(-2+Lm+4\Out+2\Rhom\right)^2}}{2m}.
\end{align}
Here we substituted the interaction parameter between $B$ and $A$ by $\chi_{AB}^*$, and truncated at $O(\Delta\epsilon)$ with  $\Delta\epsilon=\epsilon_\text{r} - \epsilon_\text{l}$.
Consistently, $\Delta \F=0$, if there is no regulator gradient ($m=0$), and when phase separation is absent ($ \epsilon_\text{l}=\epsilon_\text{r}=0,L $).
$\mathcal{G}$ depends only of the parameters of the simplified solutions (see
Eqs.~(7), main text).


\newpage

\clearpage
\subsection{Comparison of simplified and full numerical solution}

\begin{figure}[h]
\centering
\begin{tabular}{cc}
\includegraphics[width=0.5\textwidth]{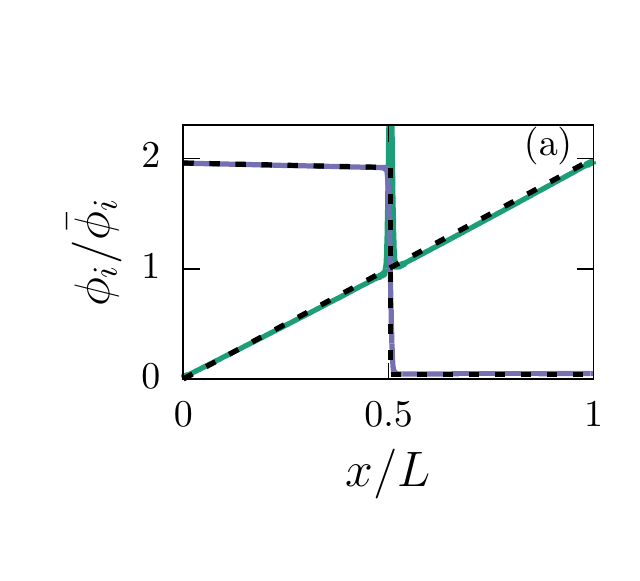} &
\includegraphics[width=0.5\textwidth]{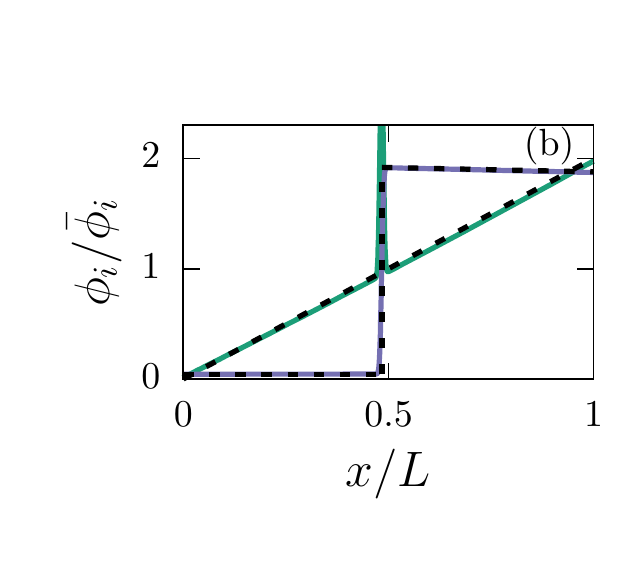} 
\end{tabular}
\caption{\label{fig:Fig_2} 
(a) Anti-correlated profile and (b) Correlated profile close to the correlated-anti-correlated transition line. The dashed black lines depict the simplified profiles (Eqs.~(7), main text) used in the analytic calculation of the free energy difference between the free energies of the two stationary solutions, $\Delta \F$.
The  peak of the regulator at the  interface between the condensed and dilute phase is neglected in the analytical ansatz.
Fixed parameters: $\chi_{AB}=4$, $\chi_{AR}=1$, $\chi_{BR}=1$, $\bar{\phi}_R=0.02$, $\phi_A=0.5$, $\kappa_R/L^2=7.63 \cdot10^{-5}$, $\kappa_A/L^2=6.10 \cdot10^{-5}$, $Ls=0.99$, $\nu=L/256$.
}
\end{figure}

\newpage

\clearpage
\subsection{Transition point is independent of the regulator gradient}

\begin{figure}[h]
\centering
\begin{tabular}{cc}
\includegraphics[width=0.5\textwidth]{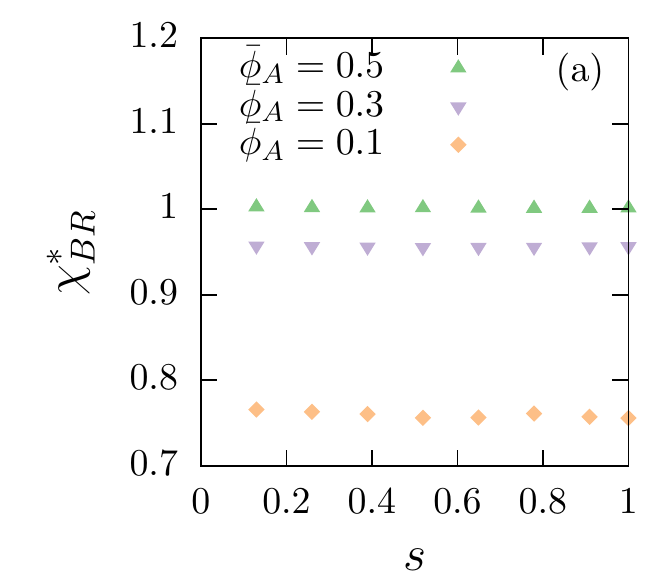} &
\includegraphics[width=0.5\textwidth]{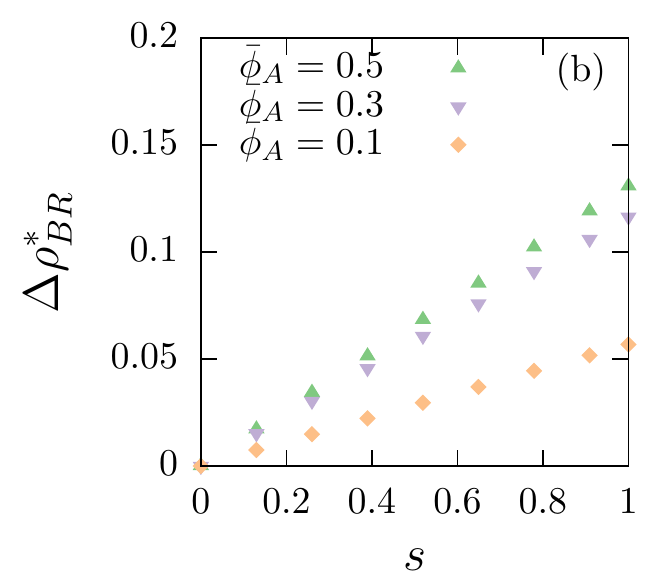} 
\end{tabular}
\caption{\label{fig:Fig_2} 
(a) The transition point is independent on the slope of the regulator gradient $s$.
(b) The jump of the order parameter at the transition point linearly increases with the slope of the gradient $s$.
The slope of this linear dependence is influenced by $\Ahom$. Fixed parameters: $\chi_{AB}=4$, $\chi_{AR}=1$, $\bar{\phi}_R=0.02$, $\kappa_R/L^2=7.63 \cdot10^{-5}$, $\kappa_A/L^2=6.10 \cdot10^{-5}$, $\kappa/L^2=6.10 \cdot10^{-5}$, $\nu=L/256$.
}
\end{figure}

\newpage

\clearpage
\subsection{Regulator Peak at the Interface}

\begin{figure}[b]
\centering
\begin{tabular}{cc}
\includegraphics[width=0.7\textwidth]{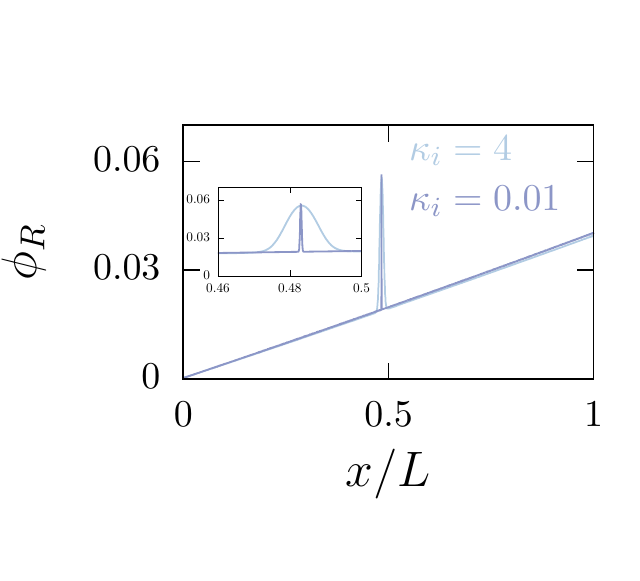}
\end{tabular}
\caption{\label{fig:PeakProfile__} 
Comparison of two regulator profiles of different $\kappa_i$ parameters.
The example $\kappa_i=4$ is very close to set of parameters we generally use, where $\kappa_A=4$, $\kappa=4$ and $\kappa_R=5$.
The case $\kappa_i=0.01$ is used as an example of small $\kappa_i$ parameters.
The plot shows that the peak area decreases if smaller $\kappa_i$ are used.
The smaller peak area is caused by a reduced peak width while the peak height is constant in good approximation.
Fixed parameters: $\chi_{AB}=4$, $\chi_{AR}=1$, $\bar{\phi}_R=0.02$, $\bar{\phi}_A=0.5$, $s=0.99$, $\nu=L/256$.\\
}
\end{figure}

The numerically obtained regulator profiles show a significant peak at the interface between the A-rich and the B-rich phase (see main text, Fig.~1(a,b)).
The emergence of the regulator peak can be understood by entropic and energetic considerations of the free energy.
For large and positive $\chi_{AR}$ and $\chi_{BR}$ (corresponding to a repulsive tendency with respect to the regulator), the energy of the system decreases as regulator accumulates at the interface.
Moreover, the entropy decreases as the composition of the interfacial region of all three components is closer to a well-mixed state.

\begin{figure}[tb]
\centering
\begin{tabular}{cc}
\includegraphics[width=0.4\textwidth]{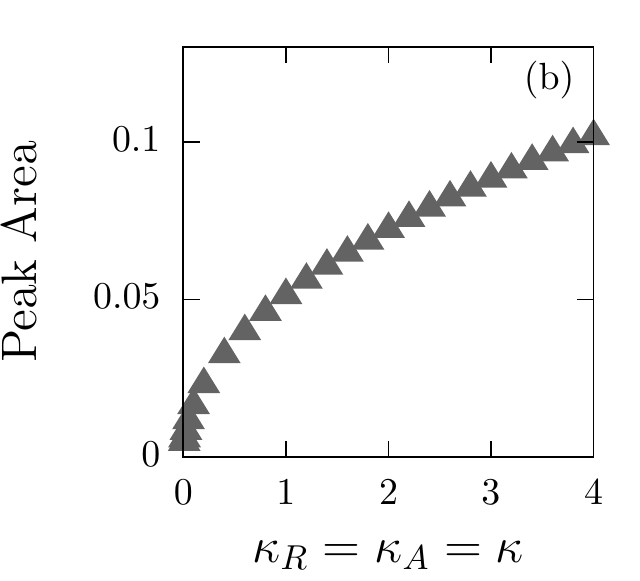}
\includegraphics[width=0.4\textwidth]{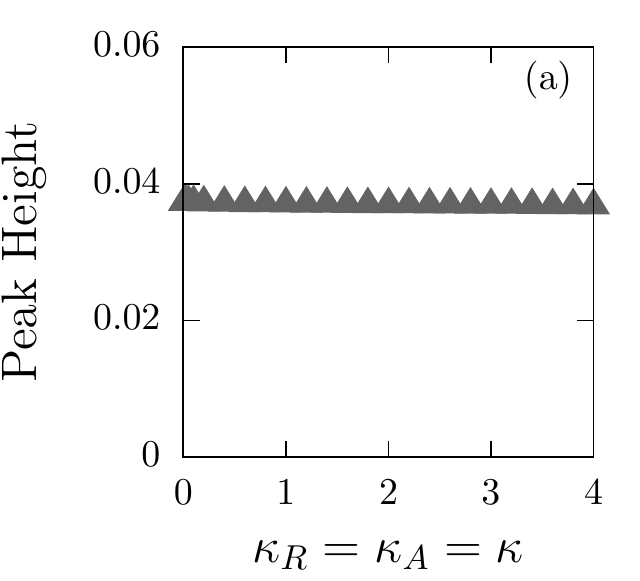}
\end{tabular}
\caption{\label{fig:PeakArea}
Peak area (left) and peak height (right) as a function of the $\kappa_i$ parameters.
These two properties are measured without the linear regulator background.
The peak height is a measure for the equilibrium value of the volume fraction of the regulator at the interface.
The peak area measures the amount of regulator material accumulated at the interface.
Here, the parameters $\kappa_A$, $\kappa_R$ and $\kappa$ are equal and changed simultaneously.
The $\kappa$ parameters have very minor influence on the peak height, it decreases only very slightly with increasing $\kappa_i$ parameters.
The influence of the $\kappa_i$ parameters on the peak area is significant.
The peak area decreases for smaller $\kappa_i$ parameters.
For very small $\kappa$ parameters, the peak area is close to zero.
Fixed parameters: $\chi_{AB}=4$, $\chi_{AR}=1$, $\bar{\phi}_R=0.02$, $\bar{\phi}_A=0.5$, $s=0.99$, $\nu=L/256$.
}
\end{figure}

The amount of regulator material that is accumulated at the interface  is strongly influenced by the $\kappa_i$-parameters; see Fig.~\ref{fig:PeakProfile__}.
In Fig.~\ref{fig:PeakArea} left, the peak area is shown for varying $\kappa_i$-parameters.
For simplicity, we chose $\kappa_A=\kappa_R=\kappa$.
The  peak area vanishes as the $\kappa_i$-parameters approach zero. This behavior is expected since these parameters set the size of the interface between the phase separated phases. 
In this limit, the estimates for the phase boundaries based on the approximate solution (main text, Eq.~(6)) are valid.

\begin{figure}[h]
\centering
\begin{tabular}{cc}
\includegraphics[width=0.4\textwidth]{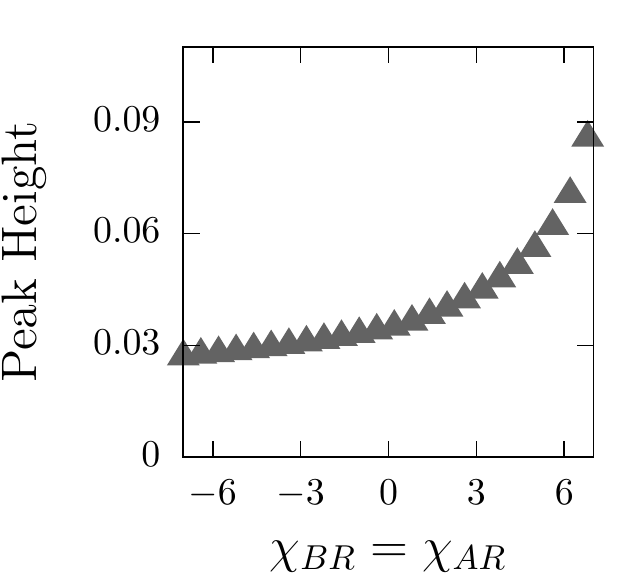}
\end{tabular}
\caption{\label{fig:PeakProfile} 
Peak height for different Flory-Huggins parameters $\chi_{AR}=\chi_{BR}$.
The peak height shows a monotonic growth for increasing $\chi_{AR}=\chi_{BR}$.
The volume fraction of the regulator at the interface growth, if the energetic interaction of the regulator with the other components becomes more repulsive.
Fixed parameters: $A=0.0077$, $\chi_{AB}=4$, $\bar{\phi}_R=0.02$, $\bar{\phi}_A=0.5$, $\kappa_R/L^2=7.63 \cdot10^{-5}$, $\kappa_A/L^2=6.10 \cdot10^{-5}$, $\kappa/L^2=6.10 \cdot10^{-5}$, $\nu=L/256$.
}
\end{figure}

However, the peak height and thereby the existence of the peak is approximately independent of $\kappa_i$ (Fig.~\ref{fig:PeakArea} right).
This indicates that the existence of the peak may depend on the interaction parameters for example. Since we also observed that the peak is more pronounced at the transition line between anti-correlated state and correlated state, we investigated the energetic influence on the peak height along the transition line.
As derived in the main text, the transition line is governed by  the condition $\chi_{AR}=\chi_{BR}$
for $\bar{\phi}_A=0.5$.
We find that the peak height increases  as a function of  the energetic parameters $\chi_{AR}=\chi_{BR}$ (Fig.~\ref{fig:PeakProfile}).
Large and positive values of $\chi_{AR}$ and $\chi_{BR}$ correspond to a repulsive tendency with respect to the regulator. 
This indicates that the energetic contribution to the free energy decreases as regulator accumulates at the interface.

\newpage
\cleardoublepage


%

\end{document}